\documentclass[aps,prx,superscriptaddress,twocolumn,eprint]{revtex4-2}
\usepackage{amsmath}
\usepackage{amsthm}
\usepackage{mathtools}
\usepackage{hyperref}
\usepackage{amssymb}
\usepackage{physics}
\usepackage{graphicx}
\graphicspath{ {figures/} }
\usepackage[caption=false]{subfig}
\usepackage{bm}
\usepackage[dvipsnames]{xcolor}
\usepackage{float}

\usepackage{comment}

\usepackage{setspace} % allow double space

\usepackage[section]{placeins} % should prevent floats from being moved to different section

\usepackage{tikz}
\usetikzlibrary{positioning, calc}

\newcommand{\p}[2]{\ensuremath{\frac{\partial #1}{\partial #2}}} %used to write partial derivatives
\newcommand{\beq}{\begin{equation}}
\newcommand{\eeq}{\end{equation}}

\begin{document}

	\title{Statics and diffusive dynamics of surfaces driven by $p$-atic topological defects}
	\author{Farzan Vafa}
	\affiliation{Center of Mathematical Sciences and Applications, Harvard University, Cambridge, MA 02138, USA}
	\author{L. Mahadevan}
	\affiliation{School of Engineering and Applied Sciences, Harvard University, Cambridge, MA 02138, USA}
	\affiliation{Departments of Physics, and Organismic and Evolutionary Biology, Harvard University, Cambridge, MA 02138, USA}
	% \pmb out how to write date
	\date{\today}

	\begin{abstract}
		
Inspired by epithelial morphogenesis, we consider a minimal model for the shaping of a surface driven by $p$-atic topological defects. We show that a positive (negative) defect can dynamically generate a (hyperbolic) cone whose shape evolves diffusively, and predict that a defect of charge $+1/p$ leads to a final semi-cone angle $\beta$ which satisfies the inequality $\sin\beta \ge 1 - \frac{1}{p} + \frac{1}{2p^2}$.  By exploiting the fact that for axisymmetric surfaces, the extrinsic geometry is tightly coupled to the intrinsic geometry, we further show that the resulting stationary shape of a membrane with negligible bending modulus and embedded polar order is a deformed lemon with two defects at antipodal points. Finally, we close by pointing out that our results may be relevant beyond epithelial morphogenesis in such contexts as shape transitions in macroscopic closed spheroidal surfaces such as pollen grains.
		
	\end{abstract}
	
	\maketitle
 
		\section{Introduction}
	\label{sec:introduction}
 
A two-dimensional surface embedded in $\mathbb R^3$ is fully described, up to rigid motions, by the first and second fundamental forms, or equivalently, the induced metric and the curvature tensor. The first fundamental form encodes the intrinsic geometry, whereas the second fundamental forms encodes both the intrinsic and extrinsic aspects of the geometry. More specifically, the eigenvalues of the second fundamental form are the two principal curvatures of the surface; their average, the mean (extrinsic) curvature, describes how the surface is embedded in $\mathbb R^3$, whereas their product, the Gaussian (intrinsic) curvature, is   independent of the embedding. For example, a cylinder and cone have zero Gaussian curvature, but non-zero mean curvature, whereas minimal surfaces, such as helicoids and catenoid, have non-zero Gaussian curvature but zero mean curvature. The six quantities characterizing the first and second fundamental forms are not all independent; for a surface to be embeddable in three dimensions, there are additional three compatibility relations (See Ref.~\cite{deserno2015fluid} for a comprehensive review of these ideas.) 

In biology, epithelial morphogenesis of thin sheet-like structures in plants and animals is responsible for the vast majority of functional structures that make up organs and organisms. These may be modeled effectively as two-dimensional surfaces whose geometry is driven by active processes that are intimately connected to the presence of orientational order in the tangent plane that modifies the embedding and in turn is modified by it. The nature of in-plane order is akin to that of polar molecules, liquid crystals, etc., or more generally to $p$-fold rotational order, denoted as ``$p$-atics''. 
There is a growing body of evidence suggesting that topological defects, singular disruptions of the rotational order, play a crucial role in guiding or controlling morphogenesis, as seen in experimental observations of cell extrusion and apoptosis~\cite{saw2017topological}, mound formation~\cite{kawaguchi2017topological,blanch2021quantifying}, layer formation~\cite{copenhagen2020topological}, and body shaping using bulges, pits and tentacles~\cite{maroudas2020topological}. Previous work on the role of defects in deformable surfaces has focused on the dynamics driven by the extrinsic geometry~\cite{seung1988defects,park1996topological,deem1996free,frank2008defects,giomi2012hyperbolic,metselaar2019topology,hoffmann2021defect} (see Ref.~\cite{al-izzi2021active} for a recent review). In contrast, in this work, following the formalism introduced in Ref.~\cite{vafa2021active}, and taking advantage of the results of Ref.~\cite{vafa2022defect}, we focus on viewing the intrinsic geometry as the fundamental field and study its dynamics.  However, unlike our previous work~\cite{vafa2021active}, where we included the effect of activity, here we consider a passive system, where there is no activity, and demonstrate that even in this passive setting the dynamics is rich. It has been known that defects drive the geometry (see for example \cite{deem1996free,frank2008defects,giomi2012hyperbolic}). What is novel here is that we find a simple and robust link between topological defects and the resulting geometry. 
	
This paper is organized as follows. We begin in Sec.~\ref{sec:model} by reviewing a minimal model for a $p$-atic on a curved surface that incorporates intrinsic geometry and then extend it to include extrinsic geometry as well. Throughout the paper, we consider the following three examples: isolated positive defect, isolated negative defect, and multiple defects.  In Sec.~\ref{sec:dynamicsIntrinsic}, we introduce the dynamical equation for intrinsic geometry, and then in Sec.~\ref{sec:dynamicsIntrinsicPlane}, we study the dynamics of intrinsic geometry of defects on the  plane. In particular, we show that a positive (negative) defect can dynamically generate a (hyperbolic) cone, and predict its shape. In Sec.~\ref{sec:embedding}, we turn to the dynamics of extrinsic geometry for axisymmetric surfaces. For an isolated positive defect, we analytically find the height $h(t)$ of the surface as a function of time $t$, and show that $h(t) \propto \sqrt{t}$. In Sec.~\ref{sec:closed}, we consider surfaces and focus on the intrinsic and extrinsic dynamics of a sphere and lemon geometry. In Sec.~\ref{sec:Willmore}, we incorporate the effect of mean curvature through the bending energy. We review the crucial fact that for axisymmetric surfaces, the intrinsic geometry entirely encodes the extrinsic geometry, which we exploit to write the bending energy in terms of the intrinsic metric. Numerically, we find that for small bending modulus, the final geometry configuration is a deformed lemon. Moreover, we propose a model for pollen grains where the transition between spherical and lemon geometries is driven by an order-disorder phase transition depending on the hydration. 
 We conclude in Sec.~\ref{sec:discussion} by reviewing our results and suggesting future directions of research.
	
	\section{Minimal model}
	\label{sec:model}
	
In this section, we first review aspects of~\cite{vafa2022defect} which develops techniques to study $p$-atic liquid crystals deep in the ordered limit on fixed curved surfaces and then apply it to  a minimal model of morphogenesis~\cite{vafa2021active}.
		
	\subsection{Isothermal coordinates}
		
Following Gauss' work~\cite{gauss1822on}, we  learn that in two dimensions it is always possible to choose local coordinates $z$ and $\bar z$, known as isothermal (conformal) coordinates, such that the metric takes the form
	\beq ds^2 = g_{z\bar z} dz d\bar z + g_{\bar z z}  d\bar z dz = 2g_{z\bar z} |dz|^2 \equiv e^{\varphi} |dz|^2 \label{eq:metric} .\eeq
	
	In terms of $z = x + iy$ and $\bar z = x - iy$, we also have
	$$ ds^2 = e^{\varphi(x,y)}(dx^2 + dy^2) .$$
	
	We thus immediately see that the metric is conformally flat, i.e. proportional to the identity matrix, where $e^{\varphi}$, known as the conformal factor, describes position-dependent isotropic stretching. Following Ref.~\cite{vafa2022defect}, in analogy to electrostatics, we will interpret $\varphi$ as the geometric potential.
	
	\subsection{Orientational order}
	
	\subsubsection{$p$-atic tensor order parameter}
	
	Now suppose our curved 2D surface is equipped with $p$-atic order, that is, $p$-fold rotational symmetry.
	Let $\mathbf{Q}$ be the $p$-atic tensor order parameter, a traceless real symmetrized rank-$p$ tensor. In terms of isothermal coordinates, since $\mathbf{Q}$ is traceless (contraction of any pair of indices vanishes), the only non-zero components of $\mathbf{Q}$ are $Q \equiv Q^{z\ldots z}$ and $\bar Q \equiv Q^{\bar z \ldots \bar z}$, where here ellipses denote $p$ copies. Also, by reality, $Q = (\bar Q)^*$. For ease of notation, let $\nabla \equiv \nabla_z$ denote the covariant derivative with respect to $z$ and $\bar\nabla \equiv \nabla_{\bar z}$ denote the covariant derivative with respect to $\bar z$. Explicitly, covariant derivatives of the $p$-atic tensor are
 \begin{subequations}
	\begin{gather}
	    \nabla Q = \partial Q + p (\partial \varphi) Q, \qquad \bar \nabla Q = \bar \partial Q\\
	\bar\nabla \bar Q = \bar \partial \bar Q + p (\bar\partial \varphi) \bar Q, \qquad \nabla \bar Q = \partial \bar Q,
	\end{gather}
 \label{eq:gradQ}
  \end{subequations}
    where partial derivatives $\partial \equiv \partial_z$ and $\bar\partial \equiv \partial_{\bar z}$.
		
	\subsubsection{Topological defects}

		\begin{figure*}
	    \centering
	    	   \subfloat[$p=2$]
	   {\includegraphics[height=3.4cm]{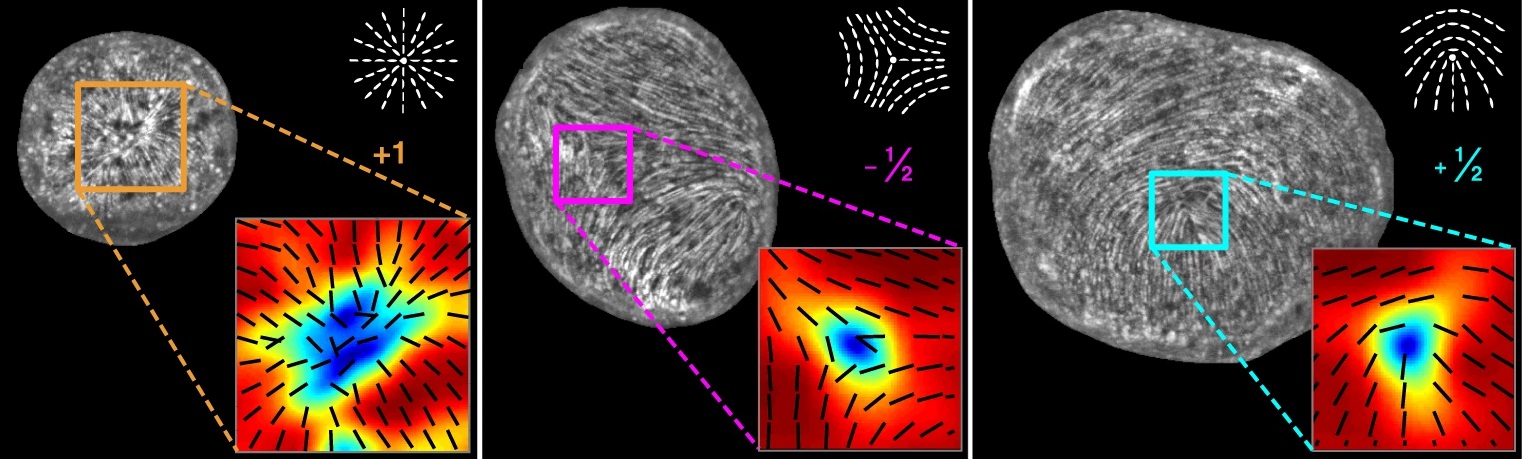}}
	    	   \subfloat[$p=6$]
	   {\includegraphics[height=3.4cm]{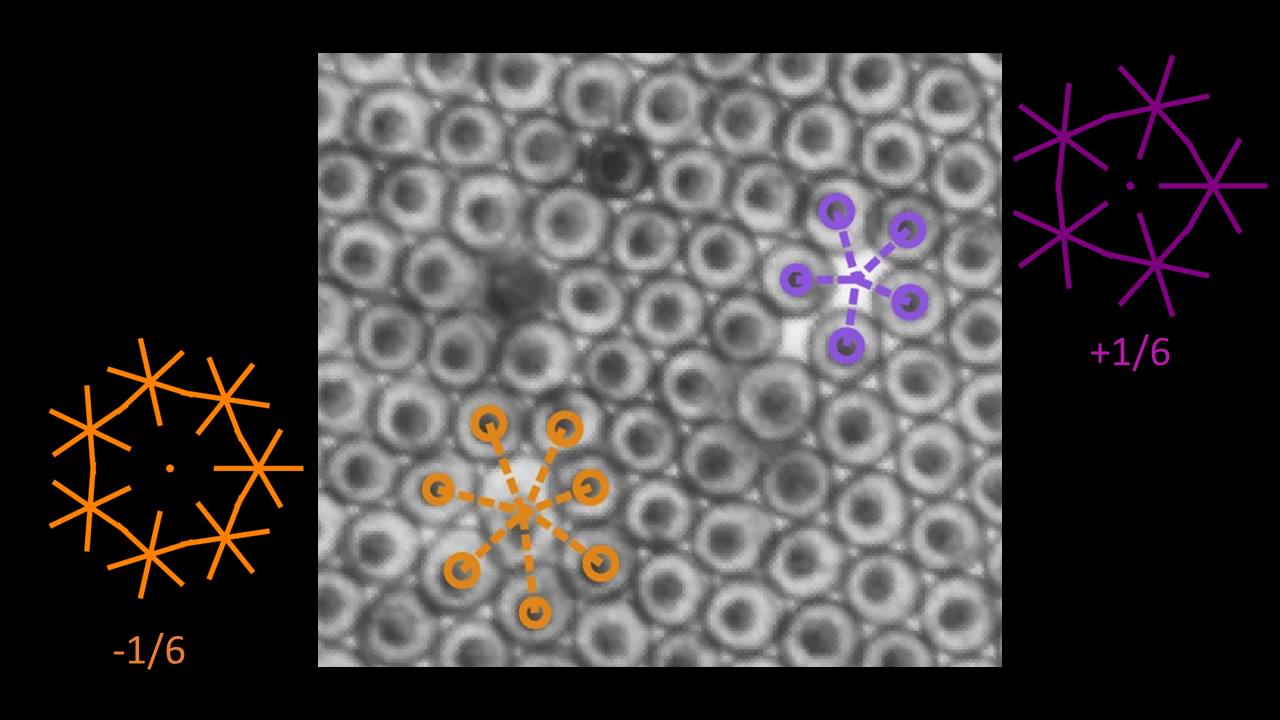}}
	    \caption{Examples of $p$-atic liquid crystals which exhibit topological defects. (a) \emph{Hydra}, adapted from Fig. 1 of ~\cite{maroudas2020topological}. Schematics in left and right corners depict textures of $+1$ and $\pm 1/2$ defects. Insets: zoomed in pictures of corresponding actin fiber orientation and scalar order parameter.  (b) Starfish embryos, adapted from Fig. 1 of \cite{tan2022odd}. Schematics in left and right corners depict textures of $\pm 1/6$ defects.}
	    \label{fig:expts}
	\end{figure*}

The in-plane orientational order can be interrupted by topological defects, where the phase of the order parameter winds around a closed loop and the amplitude vanishes. Topological defects have been observed to play a key role in diverse biological processes~\cite{saw2017topological,kawaguchi2017topological,blanch2021quantifying,copenhagen2020topological,maroudas2020topological}, in Fig.~\ref{fig:expts}, we show a few examples and sketches of defects, which we aim to describe.

	\subsection{Free energy}
	
For a surface with intrinsic in-plane $p$-atic order that is embedded in three dimensions, the three main contributions to the free energy that we consider are: (i) $\mathcal F_Q$, from the $p$-atic tensor $\mathbf{Q}$ describing features in the plane (ii) $\mathcal F_g$, from the metric $g_{ab}$ (iii) $\mathcal F_{el}$, due to the embedding. Then the total free energy $\mathcal F$ is the sum of contributions from the $p$-atic field, the intrinsic metric, and the embedding, with 
	\beq \mathcal F = \mathcal F_Q + \mathcal F_g + \mathcal F_{el} \label{eq:FGen}. \eeq
	
	In isothermal coordinates, $\mathcal F_Q$, the contribution from the $p$-atic field to the free energy (Eq.~\eqref{eq:FGen}), is given by
	\beq\mathcal F_Q = \frac{2^{p+1}}{p^2} \int d^2z \sqrt{g}[K |\nabla Q|^2 + K' |\bar \nabla Q|^2 + \epsilon^{-2} (1 - 2^p|Q|^2)^2],\label{eq:FQGen} \eeq
	where
	$$ |\nabla Q|^2 = g_{z \bar z}^{p-1}\nabla Q \bar\nabla \bar{Q}, \, |\bar \nabla Q|^2 = g_{z \bar z}^{p-1}\bar\nabla Q \nabla \bar{Q}, \, |Q|^2 = g_{z \bar z}^p Q \bar Q. $$
	Here $K,K' >0$ are Frank elastic type terms (having the same effect in flat space), and the last term governs the $p$-atic order, with $\epsilon$ controlling the microscopic $p$-atic coherence length (or defect core radius) $\xi = \sqrt{K + K'}\epsilon$.
	
	$\mathcal F_g$, the geometric contribution to the free energy (Eq.~\eqref{eq:FGen}), is written in isothermal coordinates as
	\beq \mathcal F_g = \int d^2z \sqrt{g}[2K_\varphi R\varphi + \lambda] , \label{eq:Fg}\eeq
	where $R=-2e^{-\varphi}\partial\bar\partial\varphi$ is the Gaussian curvature. $K_\varphi$ is an elastic constant penalizing changes in the curvature, and this term is a manifestation of the well-known trace anomaly, where the response of the system to conformal rescaling of the metric is proportional to the curvature~\cite{polyakov1981quantum}. $\lambda(t)$ controls the growth rate of the area. In general, $\lambda = \lambda(x,t)$, but for simplicity we will take $\lambda=\lambda(t)$, with $\lambda$ chosen such that the surface area $A = \int d^2z \sqrt{g}$ does not change. Here we will mostly focus on the case $\lambda<0$, which corresponds to positive Gaussian curvature.
	
	The final contribution to the free energy (Eq.~\eqref{eq:FGen}) is the bending energy,
	$$ \mathcal F_{el} = B\int d^2z \sqrt{g} H^2 ,$$
	where $H$ is the mean curvature~\cite{willmore1965note, helfrich1973elastic}. In Sec.~\ref{sec:Willmore}, we express $H$ in terms of $\varphi$.
	
	\subsection{Strongly ordered limit}
	
	We work deep in the ordered limit ($\epsilon \ll 1$). In this limit, $\mathcal F_Q$ (Eq.~\eqref{eq:FQGen}) is minimized when
	\beq 2^p|Q|^2 = 1 \label{eq:norm}.\eeq
	From Eq.~\eqref{eq:norm}, writing the order parameter $Q$ in terms of its amplitude $A$ and phase $\theta$ as 
    \beq Q^{z\ldots z} = A^{z\ldots z}e^{i\theta} = A e^{i\theta} \label{eq:Q}\eeq
    leads to
 	\beq A = e^{-\frac{p}{2}\varphi} \label{eq:A},\eeq
    where we have used $g_{z\bar z} = \frac{1}{2} e^{\varphi}$ from Eq.~\eqref{eq:metric}.
	Upon substitution of $Q$ (Eq.~\eqref{eq:Q} with the amplitude $A$ given by Eq.~\eqref{eq:A}) into Eq.~\eqref{eq:FQGen}, $\mathcal F_Q$ simplifies to
	\beq \mathcal F_Q = (K+K')\int d^2z \left|\left(\frac{p}{2}\right)\partial\varphi + i \partial\theta\right|^2 \label{eq:FQSimple},\eeq
	where we have used
	\begin{align*}
	\nabla_z Q^{z\ldots z} &= \left(\frac{p}{2}\partial\varphi + i\partial\theta\right)Q^{z \ldots z}\\
	\bar\nabla_{\bar z} Q^{z\ldots z} &= \left(-\frac{p}{2}\bar\partial\varphi + i\bar\partial\theta\right)Q^{z \ldots z},
    \end{align*}
    which itself was obtained by evaluating the covariant derivatives (Eq.~\eqref{eq:gradQ}) using Eqs.~\eqref{eq:Q} and \eqref{eq:A}.
    Minimizing $\mathcal F_Q$ (Eq.~\eqref{eq:FQSimple}) with respect to $\theta$ gives
    \beq \partial\bar\partial \theta = 0 \label{eq:LaplaceEq} \; .\eeq
    In the presence of a topological defect of charge $\sigma \in \mathbb{Z}/p$, the phase $\theta$ will wind by $2\pi p \sigma$. Thus a solution to Eq.~\eqref{eq:LaplaceEq} with defects $j$ at $z_j$ with charge $\sigma_j$ is
    \beq \theta = -\frac{i}{2}\sum_j (p\sigma_j) \ln \frac{z - z_j}{\bar z - \overline{z_j}} . \label{eq:theta}\eeq
	Using Eq.~\eqref{eq:theta} and the Green's function $G(z,z')$, which satisfies
	$$ \partial\bar\partial G(z,z') = \frac{1}{4}\delta^2(z - z'),$$
	we can compute the contribution of defects to $\mathcal F_Q$ (Eq.~\eqref{eq:FQSimple}), leading to
	\begin{align} \mathcal F_Q &=  2(K+K') \times \left[ -4\sum_{m \neq n} \sigma_m \sigma_n G(z_m,z_n) \right. \nonumber \\
	&\left. -\pi \sum_m \left(\sigma_m - \frac{1}{2}\sigma_m^2\right)\varphi(z_m) + \frac{1}{2}\int d^2z |\partial\varphi|^2\right] \label{eq:masterF}
	\end{align}
	(see ~\cite{vafa2022defect} for more details).
	
	The first term in Eq.~\eqref{eq:masterF} is the usual elastic interaction between defect pairs and the second term is the interaction between topological defects and the geometry~\cite{vitelli2004anomalous,vafa2022defect}, where a topological defect of charge $\sigma_m$ acquires an effective charge of $q_{m} = \sigma_m - \frac{1}{2}\sigma_m^2$. The third term is an elastic contribution to the free energy from the geometry.
	
	\section{Relaxational dynamics of the intrinsic geometry}
	\label{sec:dynamicsIntrinsic}
 
	In this paper, we are interested in the interaction between topological defects and geometry, which is captured by the last two terms of $\mathcal F_Q$ (Eq.~\eqref{eq:masterF}), which we will focus on. For simplicity, we focus on the case when the defects are frozen, i.e. we fix $Q$ and assume that the geometry responds to the presence of the defects. This assumption is valid if the defects are already at equilibrium positions, or we are in a regime where defect dynamics are slow compared to the changes in the geometry. We will start by limiting ourselves to the study of the dynamics of the \emph{intrinsic} geometry, as it is simpler but still capable of yielding insights into the shape of the surface. We will then  incorporate the \emph{extrinsic} geometry via the embedding and the mean curvature in axisymmetric cases, noting that in these situations, there is a tight link between intrinisic and extrinsic geometry.
 
    With these assumptions, the relevant part of the free energy (using the last line of Eq.~\eqref{eq:masterF} and Eq.~\eqref{eq:Fg}) is given by
	\begin{align} \mathcal F &=  -2\pi(K+K') \sum_m \left(\sigma_m - \frac{1}{2}\sigma_m^2\right)\varphi(z_m) \nonumber \\
	&\quad +\left(K + K' + 2K_\varphi\right)\int d^2z |\partial\varphi|^2 + 
	\int d^2z \sqrt{g} \lambda.
	\end{align}	
	We assume relaxational dynamics for $\varphi$, i.e.,
	\beq \partial_t \varphi = - \gamma_\varphi^{-1}\frac{1}{\sqrt{g}}\p{\mathcal F}{\varphi} \label{eq:relaxational} .\eeq
		
		Stationary solutions to Eq.~\eqref{eq:relaxational} satisfy $\p{\mathcal F}{\varphi} = 0$, leading to
	\beq \frac{1}{D}\partial_t\varphi =  -2(R - R_0) + 4\pi e^{-\varphi}\sum_j \chi_j \delta^2(z - z_j) = 0, \label{eq:stationary}\eeq
    where $D = 2\gamma_\varphi^{-1}(2K_\varphi + K + K')$ is the diffusivity, $R = -2e^{-\varphi}\partial\bar\partial\varphi$ is the Gaussian curvature, $R_0 = -2(D\gamma_\varphi)^{-1}\lambda$, and
    \beq \chi_j = \frac{K+K'}{2K_\varphi + K + K'}\left(\sigma_j - \frac{1}{2}\sigma_j^2\right) \le \sigma_j - \frac{1}{2}\sigma_j^2, \label{eq:chi_j} \eeq
	where the inequality for $\chi_j$ follows because $K,K',K_\varphi \ge 0$. We can interpret Eq.~\eqref{eq:stationary} as the Gaussian curvature $R$ is sourced by defects at $z_j$ with strengths $\chi_j$, and away from the defects, $R$ is locked to $R_0$, an effective target curvature via $R = R_0$ and thus is constant. Related aspects were noted in Ref.~\cite{giomi2012hyperbolic}.
	
	We now turn to the evolution of the geometry. Rewriting Eq.~\eqref{eq:stationary} explicitly in terms of $\varphi$, we have
	\beq  e^{\varphi}\partial_t \varphi = D\left[\partial\bar\partial\varphi + \pi \sum_{j=1}^n \chi_j\delta^2(z - z_j) + R_0 e^{\varphi}/2 \right]. \label{eq:dvarphidt}\eeq
	Eq.~\eqref{eq:dvarphidt}, except for the nonlinearity due to the $e^\varphi$ terms, looks like the regular diffusion equation with sources at positions of defects $z_j$, with strengths $\chi_j$.
	Linearizing Eq.~\eqref{eq:dvarphidt} in the neighborhood of $\varphi=0$, then we have the usual linear diffusion equation, with point sources, whose solutions can be written by convolving with the usual Green's function. The full nonlinear equation (Eq.~\eqref{eq:dvarphidt}) corresponds to Ricci flow~\cite{hamilton1982three} with sources. Moreover, the $e^\varphi$ factor gives rise to nonlinearity, which has been extensively studied by mathematicians and in fact confirms this physical intuition.
 
	We begin our analysis of Eq.~\eqref{eq:relaxational} with the analysis of defects on the plane, with the case of intrinsic geometry covered in Sec.~\ref{sec:dynamicsIntrinsicPlane} and the case of extrinsic geometry covered in Sec.~\ref{sec:embedding}. In Sec.~\ref{sec:closed}, we generalize the analysis to surfaces. We then take into account the effect of the mean curvature in Sec.~\ref{sec:Willmore}.
	
	\section{Intrinsic geometry of defects on the plane}
    \label{sec:dynamicsIntrinsicPlane}
 
	\subsection{Stationary solution}

Here we study a single defect at the origin of the plane for $R_0 = 0$. A solution to Eq.~\eqref{eq:stationary} is
	$$ \varphi = -\chi \log(z \bar z),$$
	which is in fact the geometry of a cone, i.e., the cone half angle $\beta$ satisfies $1-\sin\beta = \chi$. A positive (negative) defect thus ultimately generates a cone with positive (negative) curvature singularity. Related aspects were noted in \cite{deem1996free,frank2008defects,giomi2012hyperbolic}. We now comment that since $\chi \le \sigma - \sigma^2/2$, then we predict that there is an upper bound for $\chi$. Since $\sigma$ is in units of $1/p$, then for $p=1$, $\chi \le 1/2$, which corresponds to $\sin\beta \ge 1/2$, i.e., $\beta \ge \pi/6$. This means that there is an upper bound to how sharp a cone can be. For all $p$-atics, the upper bound is given by $p=1$, and as $p$ increases, the cone becomes less sharp. For example, we predict for a nematic that $\chi \le 1/2 - (1/2)^2/2 = 3/8$.

\subsection{Dynamics}

We study the evolution of the intrinsic geometry by starting with the case of an isolated defect on the plane $(R_0 = 0)$, with initial condition $\varphi(z,\bar z, t=0)=0$, i.e. flat geometry. To see how a defect can generate non-trivial geometry,
    we assume axisymmetric solution, i.e. $\varphi = \varphi(r,t)$, which upon substitution of $\partial\bar\partial = \frac{1}{4}\left(\frac{\partial^2}{\partial r^2} + \frac{1}{r}\p{}{r}\right)$ and $\delta^2(z) = \frac{1}{2\pi r}\delta(r)$
	into Eq.~\eqref{eq:dvarphidt}, gives 
	\beq \partial_t \varphi = De^{-\varphi}[\frac{1}{4}\frac{\partial^2\varphi}{\partial r^2} + \frac{1}{4r}\frac{\partial\varphi}{\partial r} + \frac{\epsilon}{2 r}\delta(r)] \label{eq:dvarphidtpolar}.\eeq
	We propose a self-similar ansatz,
	\beq \varphi = \varphi(u \equiv \frac{r^2}{Dt}), \label{eq:uansatz}\eeq
	 which upon substitution into Eq.~\eqref{eq:dvarphidtpolar} gives
	\beq -ue^{\varphi}\partial_u \varphi = \partial_u(u\partial_u \varphi) + \chi \delta(u)\label{eq:dvarphidu} .\eeq
	
\begin{figure}[t]
		\centering
		\includegraphics[width=\columnwidth]{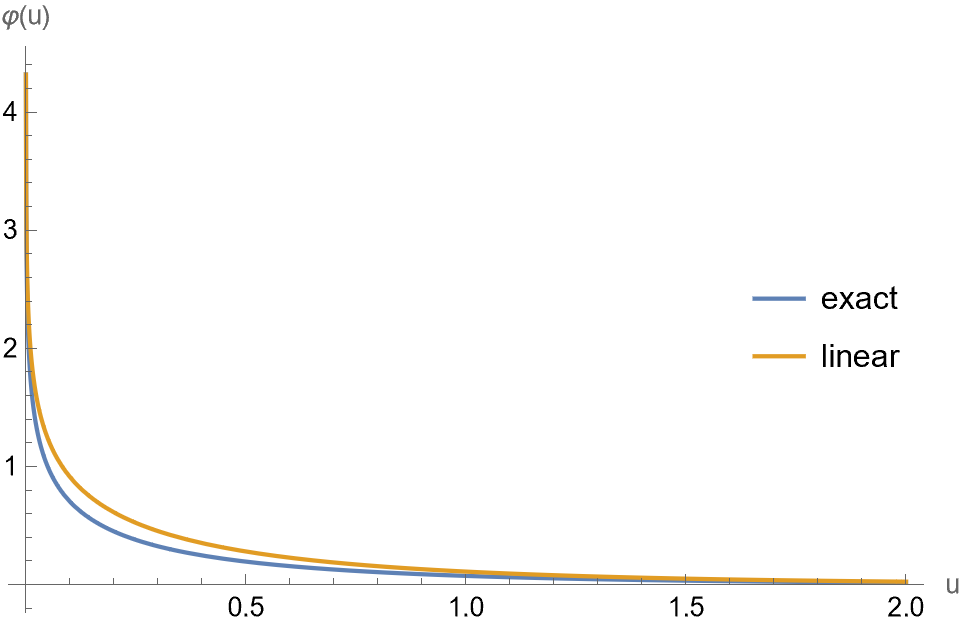}
		\caption{Plot of $\varphi(u)$ for the exact diffusion (Eq.~\eqref{eq:dvarphidu}) and linearized equation (Eq~\eqref{eq:dvarphiduLinear}) for $\chi=0.5$.}
		\label{fig:phiSoln}
	\end{figure}

	In Fig.~\ref{fig:phiSoln}, we show the comparison of the geometric diffusion equation (Eq.~\eqref{eq:dvarphidu}) with the solution of the linearized diffusion equation 
    \beq -u\partial_u \varphi = \partial_u(u\partial_u \varphi) + \chi \delta(u)\label{eq:dvarphiduLinear}\eeq
    (for $\chi = 0.5$), and find excellent agreement. The difference becomes more significant as $\chi$ approaches 1.
	
	The $u\to0$ limit of Eq.~\eqref{eq:dvarphidu} corresponds to the steady state of Eq.~\eqref{eq:dvarphidt}, and yields the short distance / long time behavior of the geometry. To understand the long distance / short time behavior, we now consider the $u\to\infty$ limit. 
	
	Starting from flat configuration $\varphi=0$, and to leading order in $\varphi$, Eq.~\eqref{eq:dvarphidu} becomes
	\beq -u\partial_u \varphi  = \partial_u (u\partial_u\varphi) \label{eq:dvarphiduLau}.\eeq
	Let $f(u) = u\partial_u\varphi$. Then integrating Eq.~\eqref{eq:dvarphiduLau} once immediately gives $f = -C_1e^{-u}$, with constant of integration $C_1>0$ (because $f=u\partial_u\varphi < 0$), and so integrating Eq.~\eqref{eq:dvarphiduLau} once more immediately gives
	\beq \varphi = -C_1\int_\infty^u \frac{du'}{u'}e^{-u'}\label{eq:GreenDiffusion} ,\eeq
	where by construction $\varphi(u=\infty) = 0$ and the constant of integration $C_1$ can in principle be determined by matching this solution with the one for $u\to0$, leading to $C_1 = \chi$. Intriguingly, in both the $u\to0$ and $u\to\infty$ limits, the exponential factor $e^{\varphi}$ was negligible. 
	In fact, Eq.~\eqref{eq:GreenDiffusion} is the Green's function of the geometric diffusion equation, ignoring the $e^\varphi$ terms.
	
	Having seen that the solution to the linearized equation is sufficiently good for sufficiently small $\chi_j$, then in this regime we can get a good approximation for our dynamical solution by simply solving the linearized equation, which leads to
	$$ \varphi(r,t) = \sum_j \chi_j \int_0^t \frac{dt'}{t'}\exp{-\frac{|r - r_j|^2}{Dt'}} .$$

    \section{Extrinsic geometry of a defect on the plane}
	\label{sec:embedding}
	
    We now find the extrinsic geometry of a defect on the plane. In particular, for a single positive defect, we find the exact dynamical solution, and show that at all times, the height grows as $h(t) \propto \sqrt{t}$. We begin by considering a surface over the flat $(z',\bar z')$ plane with the height $h(z',\bar z')$ above the plane which reproduces the intrinsic metric we have found. In other words, we look for solutions to
	\beq ds^2 = e^{\varphi} dz d\bar z = dh^2 + dz' d\bar z' \label{eq:ds2tilde}. \eeq
	In terms of polar coordinates, $z = re^{i\phi}$ and $\tilde z' = r' e^{i\phi'}$, Eq.~\eqref{eq:ds2tilde} becomes
	\beq e^{\varphi}(dr^2 + r^2 d\phi^2) = dh^2 + dr'^2 + r'^2 d\phi'^2 \label{eq:ds2tildepolar}.\eeq
    We now consider separately the cases of a single positive or negative defect.
 
	\subsection{Positive defect}
	
	For a positive defect, noting that $\phi = \phi'$ implies that
	\beq r'^2 = r^2e^{\varphi} \label{eq:rtilde},\eeq
	and so
	\beq dh^2 + dr'^2 = e^{\varphi}dr^2\label{eq:ds2polar} .\eeq
    As in the case of intrinsic geometry, it is useful to write in terms of $u \equiv r^2/(Dt)$. Dividing Eq.~\eqref{eq:ds2polar} by $dr^2$ and using
	\begin{gather*}
    \frac{dr'}{dr} =(1 + \frac{r}{2}\partial_r\varphi)e^{\varphi/2} = (1 + u\partial_u \varphi)e^{\varphi/2} \\
    \frac{dh}{dr} = \frac{dh}{du} \frac{du}{dr} = \frac{2u}{\sqrt{uDt}}
    \end{gather*}
    leads to
	\beq \frac{h}{\sqrt{Dt}} = I-\frac{1}{2}\int_0^u du \sqrt{[-2\partial_u\varphi - u(\partial_u \varphi)^2]}e^{\varphi/2}  , \label{eq:h}\eeq
	where
	$$ I = \frac{1}{2}\int_0^\infty du \sqrt{[-2\partial_u\varphi - u(\partial_u \varphi)^2]}e^{\varphi/2}$$
	is a constant that depends only on $\chi$.
	
	We now study two limits of Eq.~\eqref{eq:h}: the $u\ll 1$ and $u\to\infty$ limits.
	
	\subsubsection{$u \ll 1$ limit}
	
\begin{figure}[t]
		\centering
		\subfloat[]{\includegraphics[width=\columnwidth]{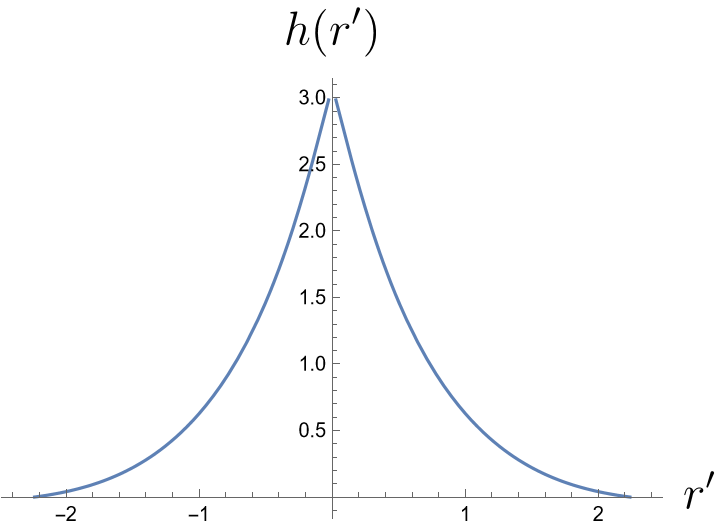}}\\
		\subfloat[]{\includegraphics[width=\columnwidth]{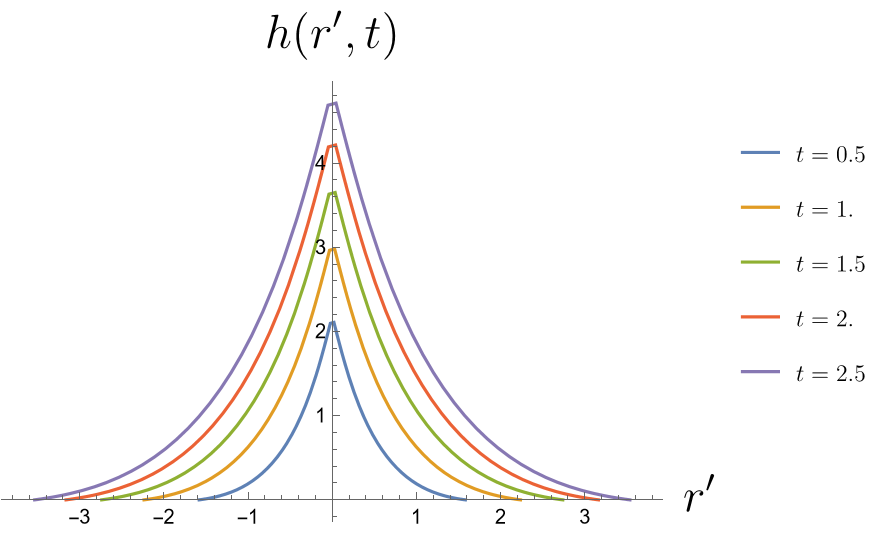}}
		\caption{Plots of $h(r',t)$ (Eq.~\eqref{eq:h}) (a): $h(r',t)$ for $t=1.0$. (b): $h(r',t)$ for $t=0.5,1.0,\ldots,2.5$, $D=1.0$, $\chi=0.5$. As $t$ increases, curve is dilated by factor of $\sqrt{t}$ (both $r'$ and $h(r',t)$). Parameters used are $D=1.0$ and $\chi=0.5$.}
		\label{fig:height}
	\end{figure}

	We first study the $u \ll 1$ limit of Eq.~\eqref{eq:h}, or equivalently, the $t\to \infty$ or $r=0$ limit. In this limit, since $\varphi = -\chi \log u$, then Eq.~\eqref{eq:h} simplifies to
	\begin{align}
		\frac{h}{\sqrt{Dt}} &\approx I -\frac{1}{2}\int_0^u du \sqrt{1 - (1-\chi)^2} u^{-\frac{\chi+1}{2}} \nonumber \\
		&= I - \frac{\sqrt{1 - (1 -\chi)^2}}{1-\chi} u^{\frac{1-\chi}{2}} \label{eq:husmall}
	\end{align}
	Then from Eq.~\eqref{eq:rtilde},
	$$ r'^2 = r^2 e^{\varphi} \implies u = \left(\frac{r'^2}{Dt}\right)^{\frac{1}{1-\chi}} ,$$
	and thus upon substitution into Eq.~\eqref{eq:husmall}, we have
	\beq h = I\sqrt{Dt} - \cot\beta \ r' \label{eq:hcone},\eeq
	where $\sin\beta = 1-\chi$. What this means is that we have a cone with half-cone angle $\beta$ where the height grows proportional to $\sqrt{t}$ with proportionality $I\sqrt{D}$ that is determined in our model. (See Fig.~\ref{fig:height} for a plot).
	
	Here for simplicity we have assumed $\lambda=0$. If we were to restore $\lambda$, then away from the defect, in the steady state we get constant Gaussian curvature, whose sign is opposite that of $\lambda$. This makes contact with Ref.~\cite{frank2008defects}, which corresponds to studying the case where $\lambda > 0$. However, the main difference is that at the defect, we predict a finite, fixed angle cone, whereas Ref.~\cite{frank2008defects} finds that the slope diverges and comment that they need mean curvature to smooth the divergence. Note that our results are consistent with the experimental observation of positive defects being correlated with positive Gaussian curvature in \emph{Hydra}~\cite{maroudas2020topological}.
	
	\subsubsection{$u\to\infty$ limit}
	
	We now study the opposite limit, $u\to\infty$ of Eq.~~\eqref{eq:h} (or equivalently, $t\to 0$). Then starting from flat configuration $\varphi=0$, we can assume $\varphi$ is small. Therefore, in the $u\to\infty$ limit (and thus to leading order in $\varphi$),	
	\begin{align*}
		\frac{h}{\sqrt{Dt}} &\approx \frac{1}{2}\int_u^\infty du \sqrt{-2\partial_u\varphi} \\
		&= \frac{1}{2}\sqrt{2C_1}\int_u^\infty du \sqrt{\frac{e^{-u}}{u}} \nonumber \\
		&= \sqrt{\pi C_1} \mathrm{erfc}(\sqrt{u/2}) ,
	\end{align*}
	where 
	$$ \mathrm{erfc}(x) = 1 - \mathrm{erf(x)} = 1 - \frac{2}{\sqrt{\pi}} \int_0^x e^{-t^2}dt$$
	is the complementary error function. See Fig.~\ref{fig:height} for plots.

	If we had more than one positive defect, then as long as the defects are not influencing each other diffusively, that is, the distance between defects $\ell \gg \sqrt{Dt}$, 
	then each defect would create its own conical geometry.
	
	\subsection{Negative defect}

	As before, we look for solutions to Eq.~\eqref{eq:ds2tildepolar}.
    Unlike the case of positive defect, here there are no rotationally symmetric embeddings--even though the intrinsic metric is rotationally invariant, the embedding breaks the rotational symmetry. Hence we will assume that $\varphi = \varphi(r)$, $h = f(r')\cos(m\phi')$, $r' = r'(r,\phi)$, and $\phi' = \phi'(\phi)$. For example, $m=2$ corresponds to a regular saddle, and $m=3$ corresponds to a monkey saddle. By equating the coefficients of the differentials in Eq.~\eqref{eq:ds2tildepolar}, we find that we have to solve the following coupled system of nonlinear equations:
	\begin{align*}
		\p{r'}{\phi} &= mf\p{f}{r'}\frac{\cos(m\phi')\sin(m\phi')\p{\phi'}{\phi}}{1+\left(\p{f}{r'}\right)^2\cos^2(m\phi')} \\
		e^{\varphi}r^2  &= \left[\frac{m^2f^2\sin^2(m\phi')}{1+\left(\p{f}{r'}\right)^2\cos^2(m\phi')} + r'^2\right]\left(\p{\phi'}{\phi}\right)^2\\
		e^{\varphi} & = \left(1 + \left(\p{f}{r'}\right)^2\cos^2(m\phi')\right) \left(\p{r'}{r}\right)^2\\
		\phi'(\phi) &= \phi'(\phi+2\pi) .
	\end{align*}
	
\begin{figure}[t]
		\centering
		\subfloat[$m=2$]
		{\includegraphics[width=.45\columnwidth]{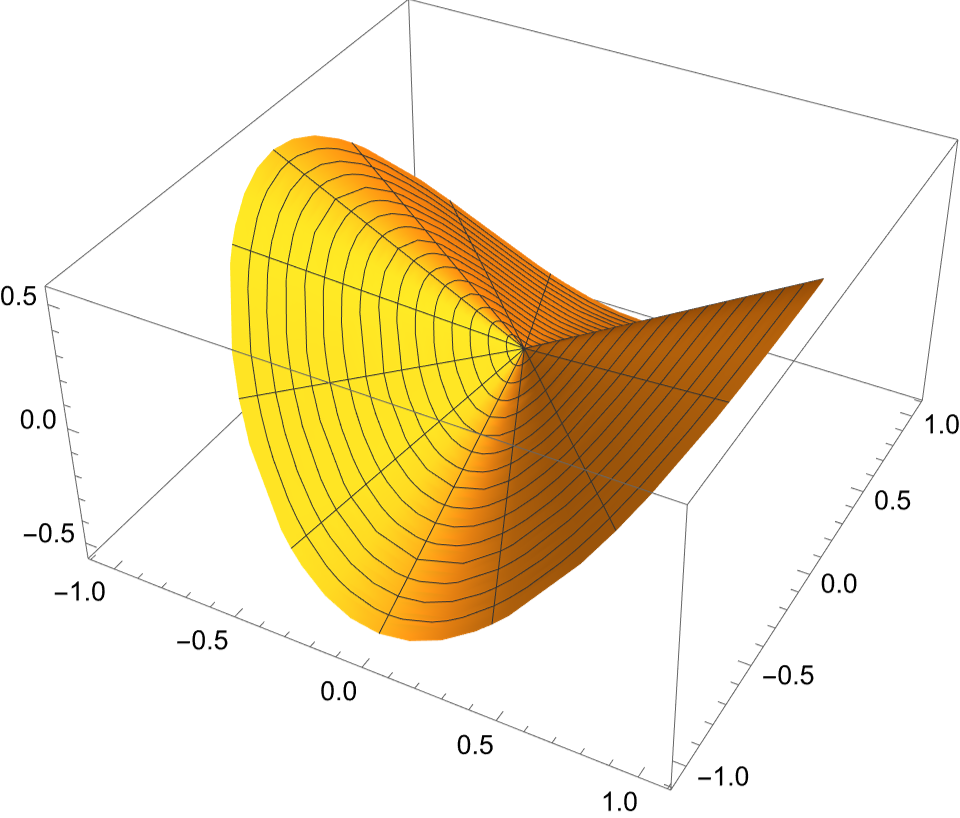}}
        \subfloat[$m=3$]
        {\includegraphics[width=.45\columnwidth]{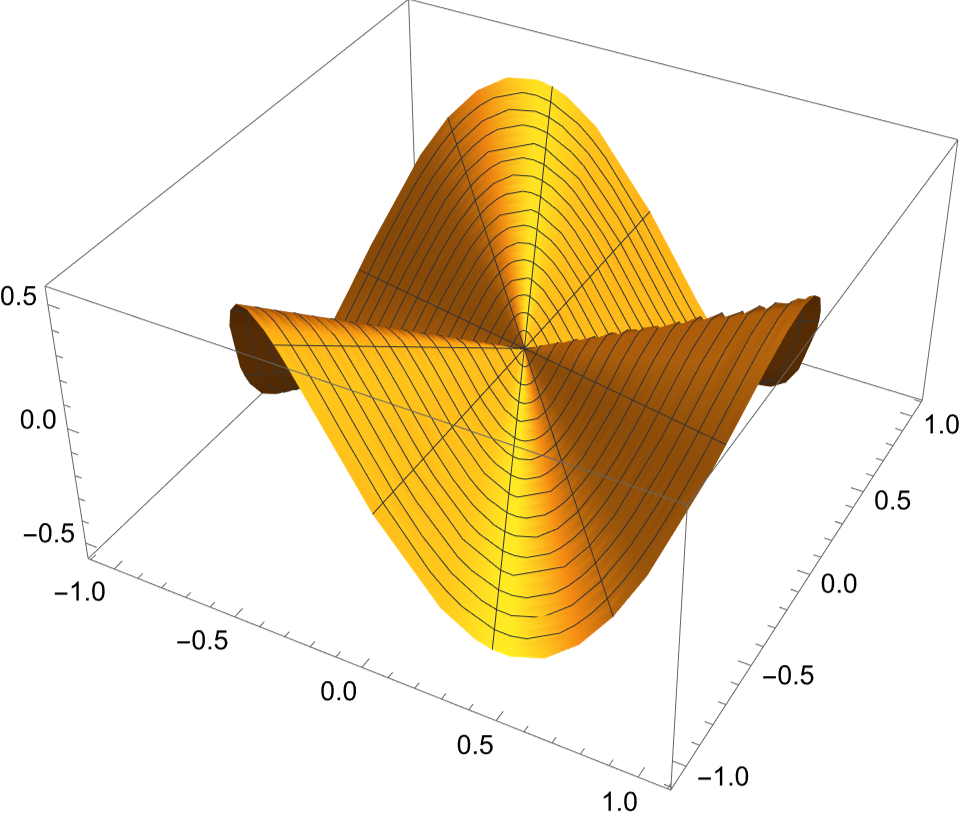}}
		\caption{Plot of saddle geometry (Eq.~\eqref{eq:saddle}) for $\chi=-0.15$ for (a) $m=2$, i.e. the regular saddle) (b) $m=3$, i.e. the monkey saddle).}
		\label{fig:saddle}
	\end{figure}

	We now check explicitly for a hyperbolic cone (see Fig.~\ref{fig:saddle} for a diagram). The embedding for a hyperbolic cone is
	\begin{subequations}
		\begin{align*}
			x &= r'\cos\phi' \\
			y &= r'\sin\phi' \\
			h &= a r' \cos(m\phi') .
		\end{align*}
		\label{eq:saddle}
	\end{subequations}
	
	With the following change of variables
	\begin{align*}
		r &= \left[(1+\chi)\sqrt{1 + a^2 \cos^2(m\phi')}r'\right]^{\frac{1}{1+\chi}} \\
		\phi &= \frac{1}{1+\chi}\int_0^{\phi'} d\phi' \frac{\sqrt{1 + a^2[1 + (m^2-1)\sin^2\phi]}}{1 + a^2 \cos^2(m\phi')} ,
	\end{align*}
	with $a$ satisfying
	$$ 2\pi(1+\chi) = \int_0^{2\pi}d\phi \frac{\sqrt{1 + a^2[1 + (m^2-1)\sin^2\phi]}}{1 + a^2 \cos^2\phi} ,$$
	the metric (Eq.~\eqref{eq:ds2tildepolar}) takes the form
	$$ ds^2 = dr'^2 + r'^2d\phi'^2 + dh^2 ,$$
	as expected. 
	
    \section{Defects on surfaces}
	\label{sec:closed}
	
While we can consider the general case of surfaces with constant positive or negative Gaussian curvature, by Hilbert's theorem we cannot embed constant negative Gaussian curvature surfaces in $\mathbb R^3$~\cite{hilbert1901ueber}. Ideally, we would like to answer the following questions: (i) What kinds of stationary solutions exist? (ii)  Are they stable? (iii) What is the embedding? (iv) What are the dynamics of the intrinsic and extrinsic geometry? To attempt answers to these questions, we first consider multiple defects on surfaces, and then in more detail multiple defects on the sphere.
 
\subsection{Intrinsic geometry}

We begin by reviewing the case of surfaces. Integrating Eq.~\eqref{eq:stationary} over the surface gives
	$$ \frac{1}{D} \partial_t A = 4\pi(2g-2) + 4\pi\sum_j \chi_j +  2R_0 A = 0,$$
	where $g$ is the genus of the surface and $A = \int d^2z \sqrt{g}$ is the area, leading to
	$$ (2-2g) - \sum_j \chi_j =  \frac{R_0 A}{2\pi} .$$
	Since away from the defects, the sign of the constant curvature is the same as the sign of $R_0$, from the above it follows that the sign of the LHS correlates with the sign of the constant curvature away from the defects, i.e.
	\beq (2-2g) \lesseqqgtr \sum_j \chi_j \iff R \lesseqqgtr 0\label{eq:Euler} .\eeq
	In particular, for the case of the sphere, $g=0$, in which case
	\beq 2 \lesseqqgtr\sum_j \chi_j \iff R \lesseqqgtr 0 . \label{eq:EulerSphere}\eeq
	We now comment on the stability of the stationary solution. For positive curvature, $R > 0$, the stationary solution is not stable if $R_0$ is a constant since $\partial_tA = 2 R_0 D A + const.$ which leads to runaway. However, in principle we can choose $R_0$ to be time-dependent, so that we can always ensure that the stationary solution be achieved, as we shall assume. For negative curvature, $R < 0$, the stationary solution is automatically stable. Regardless of whether we are at stationary point or not, we can always choose $R_0(t)$ such that the area does not change, as we will be assuming in the following. 
	
	We now consider in more detail two cases in turn: $R>0$ and $R<0$.
	
	\subsubsection{$R>0$}
	
 We first consider $R>0$ on the sphere. There is no stationary solution to Eq.~\eqref{eq:relaxational} for $n=1$. For $n=2$, there exists a stationary solution if and only if $\chi_1 = \chi_2$ and the defects are at antipodal points~\cite{troyanov1989metrics}. In this case, the shape resembles that of a lemon (see Fig.~\ref{fig:lemon}). For $n\ge 3$ defects and $\chi_i \in (0, 1)$, denoting the deficit angle $2\pi\chi_i$, and assuming the Troyanov inequality~\cite{troyanov1991prescribing,luo1992liouville},
	\beq 2 \max{\chi_i} < \sum_i \chi_i < 2 \label{eq:Troyanov} ,\eeq 
	then there exists a $2n-6$ parameter family of metrics on the sphere with constant positive scalar curvature with $n$ conical singularities given by deficit angles $2\pi\chi_i$. Here, the $2n$ counts the coordinate degree of freedom, but there are $6$ constraints associated with the M\"{o}bius transformations of the sphere. The first inequality is an intriguing prediction which would be interesting to interpret physically, while the second inequality in Eq.~\eqref{eq:Troyanov} follows from Eq.~\eqref{eq:EulerSphere}. 
	
	For $p$-atic defects on a sphere that we discuss later, all the $\chi_i$ are positive and equal, and so this inequality is automatically satisfied. Moreover, as $t\to\infty$, in all cases where a solution exists, the solution indeed converges to the unique constant curvature solution~\cite{yin2010ricci,mazzeo2015ricci,phong2020the}, and can be embedded uniquely in $\mathbb R^3$ up to translation and rotation~\cite{galvez2013surfaces}. We note that even though our model does not directly contain any information about the embedding of the surface in $\mathbb R^3$ or the extrinsic geometry, the solution leads to recovering unique extrinsic geometry! Moreover, Ref.~\cite{mondello2015spherical} extended the existence of stationary solutions to allow some $\chi_i <0$, and found that a solution can exist if additional constraints on $\chi_i$ are satisfied. Note that naively, if we had defects of mixed sign, then we would expect them to annihilate each other due to the Coulomb interaction, and is consistent with the mathematical result that the steady state metric isn't unique~\cite{troyanov1991prescribing}.
	
	\subsubsection{$R<0$}
	
	We now consider $R<0$. In this case, for a compact Riemann surface $S$ of genus $g$ with all $\chi_i < 1$, we know from Eq.~\eqref{eq:Euler} that we need
	$$ \sum_i \chi_i > 2-2g ,$$ 
	which has also been shown to be sufficient~\cite{troyanov1991prescribing}. Note that here we are not assuming that all $\chi_i$ are positive. The position degree of freedom gives $2n$ degrees for genus $g>1$, $2n-2$ for $g=1$ ($-2$ accounts for two translations of torus), and $2n-6$ for $g=0$ ($-6$ accounts for M\"{o}bius transformations of the sphere).	
	
	\subsubsection{$p$-atic on the sphere}
	
	Here we consider in more detail a $p$-atic on the sphere. Since the net charge is 2, we consider $2p$ defects each of the minimal charge $+1/p$. Since according to Eq.~\eqref{eq:chi_j}, $\sum_j \chi_j < 2$, then $R$ is constant positive away from the defects because of Eq.~\eqref{eq:EulerSphere}. Moreover, the LHS of the Troyanov inequality Eq.~\eqref{eq:Troyanov} is also satisfied because all the $\chi_j$ are equal. Thus a unique solution exists.
	
	An example where we can explicitly write the stationary state metric is for $p=1$, corresponding to polar liquid crystal, with two $+1$ defects on the sphere with equal deficit angles of $\chi$, which we place on the north and south poles of the sphere. We construct this metric by starting from the
	the round metric (spherically symmetric metric),
	$$ ds^2 = 4\frac{1}{(1 + |z|^{2})^2} |dz|^2 ,$$
	which has Gaussian curvature of 1. The coordinate transformation $z\to z^{1-\chi}$ then gives rise to conical singularities of strength $\chi$ at the north and south poles, giving the round conical metric~\cite{troyanov1989metrics},
	\beq ds^2 = 4(1-\chi)^2\frac{|z|^{-2\chi}}{(1 + |z|^{2(1-\chi)})^2} |dz|^2 . \label{eq:conical}\eeq
	Here the Gaussian curvature is still 1 (away from the poles). At the two poles $z=0$ and $z=\infty$, there are conic singularities of strength $\chi$. We call this the lemon geometry.
	
	\subsection{Extrinsic geometry}
	
	For the lemon geometry, defined by Eq.~\eqref{eq:conical}, the embedding $x_i$ is~\cite{carmo2016differential}
	
	\begin{subequations} 
		\begin{align*}
			x_1 &= a \sin\theta \cos\phi \\ 
			x_2 &= a \sin\theta \sin\phi \\
			x_3 &= \int_0^\theta d\theta' \sqrt{1 - a^2\cos^2\theta'} ,
		\end{align*}
		\label{eq:lemon}
	\end{subequations}
	where $\theta \in [0,\pi]$, $\phi \in [0,2\pi]$, and $a = 1-\chi$.
	
	It can be checked that
	$$ ds^2 = dx_1^2 + dx_2^2 + dx_3^2 = d\theta^2 + a^2\sin^2\theta d\phi^2 = e^{\varphi}|dz|^2 , $$
	where
	\begin{align*} 
	z &= \left(\tan\frac{\theta}{2}\right)^{1/a} e^{i\phi} \\
	e^{\varphi} &= \frac{a^2 \sin^2\theta}{ \left(\tan\frac{\theta}{2}\right)^{2/a}} .
	\end{align*}	
	See Fig.~\ref{fig:lemon} for a plot.
	
	\section{Including mean curvature}
	\label{sec:Willmore}
	
	In this section, we consider the full model by including the effect of the bending energy. The key is that for axisymmetric surfaces, the extrinsic geometry is entirely encoded by the intrinsic geometry. We first review this fact and explicitly express the mean curvature, via the principal curvatures $\kappa_1$ and $\kappa_2$, in terms of $\varphi$.

 For an axisymmetric surface, the embedding is
	$$ X^i = (\rho(u)\cos\theta, \rho(u)\sin\theta, h(u)) ,$$
 from which follows that the metrix is
 \beq ds^2 = \left(dX^i\right)^2 = \left(\rho'^2 + h'^2\right) du^2 + \rho^2 d\theta^2 \label{eq:ds2axisymmetric}.\eeq
	The task at hand is to express the mean curvature,
	\beq H = \frac{1}{2}(\kappa_1 + \kappa_2) ,\eeq
	and thus the principal curvatures,
 \begin{subequations}     
	\begin{align}
		\kappa_1 &= \frac{1}{2}\frac{\frac{d^2h}{d\rho^2}}{\left(1 + \left(\frac{dh}{d\rho}\right)^2\right)^{3/2}}\\
		\kappa_2 &=  \frac{1}{2}\frac{\frac{dh}{d\rho}}{\rho\left(1 + \left(\frac{dh}{d\rho}\right)^2\right)^{1/2}} ,
	\end{align}
    \label{eq:kappadefn}
  \end{subequations}
	in terms of $\varphi$.
	
	Choosing the parameter $u$ in the metric (Eq~\eqref{eq:ds2axisymmetric}) such that~\cite{arteaga2007infinitesimal}
	\beq \rho'^2 + h'^2 = \rho^2 \label{eq:uchoice} \eeq
	immediately gives
    \beq ds^2 = \rho(u)^2(du^2 + d\theta^2) \label{eq:metricrho} .\eeq
	$(\rho,\theta)$	can be viewed as cylindrical coordinates, and in terms of isothermal coordinates, $z = u + i\theta$. In these coordinates,
	$$ ds^2 = e^{\varphi(z,\bar z)} |dz|^2 ,$$
	where
	\beq \rho(u)^2 = e^{\varphi(u)} \label{eq:rho}\eeq
	and $u =(z + \bar z)/2$.
	Since $\theta$ is periodic, then $z \sim z +2\pi i$. Thus rotating the coordinates by $\alpha$ corresponds to shifting $z$ by $i\alpha$. This metric (Eq.~\eqref{eq:metricrho}) is manifestly rotational invariant as $\rho$ (Eq.~\eqref{eq:rho}) depends only on $u$ and not $\theta$. We briefly explicitly consider the examples of sphere and lemon before turning to the general axisymmetric surfaces.
	
	\subsection{Sphere geometry}
	
	The round sphere has metric
	\beq ds^2 = \frac{4|dw|^2}{(1 + |w|^2)^2} \label{eq:sphere}.\eeq
	We define $w = \exp[z]$ so that the phase rotation of $w$ is identified with shift of the imaginary component of $z$. In terms of $z$, Eq.~\eqref{eq:sphere} becomes
	$$ ds^2 = \frac{4|d e^z|^2}{(1 + |e^{z}|^2)^2} = \frac{4 |e^z|^2 |dz|^2}{(1 + |e^z|^2)^2} .$$
	Using $|e^z| = e^u$, we learn that
	$$ ds^2 = \frac{4|d e^z|^2}{(1 + |e^{z}|^2)^2} = \frac{4 e^{2u} |dz|^2}{(1 + e^{2u})^2} = \sech[2](u)|dz|^2 ,$$
	and thus $\rho(u) = \sech(u)$. For the height,
	$$ h = \pm \int du \sech[2](u) = \pm \tanh(u) .$$
	Note that as a consistency check, $\rho^2 + h^2 = 1$, which indeed describes a sphere.
	
	\subsection{Lemon geometry}
	
	The lemon geometry has metric
	\beq ds^2 = 4(1-\chi)^2\frac{|w|^{-2\chi}}{(1 + |w|^{2(1-\chi)})^2} |dw|^2 \label{eq:ds2lemon}.\eeq
	In terms of $z=\ln w$, Eq.~\eqref{eq:ds2lemon} becomes
	$$ ds^2 = 4(1-\chi)^2\frac{|e^z|^{-2\chi} |d e^z|^2}{(1 + |e^{z}|^{2(1-\chi)})^2} = 4(1-\chi)^2\frac{e^{2u(1-\chi)} |dz|^2}{(1 + e^{2u(1-\chi)})^2} .$$
	Thus we learn
	$$ \rho(u) = (1-\chi)\sech(u(1-\chi)) .$$
	
	\subsection{General axisymmetric surfaces}
	
	We now turn to general axisymmetric surfaces. We first note, using Eq.~\eqref{eq:uchoice}, that
	$$ \left(\frac{dh}{d\rho}\right)^2 = \left(\frac{h'}{\rho'}\right)^2 = \frac{\rho^2 - \rho'^2}{\rho'^2} ,$$
	from which follows that
	\beq\frac{dh}{d\rho} = \sqrt{\frac{4}{\varphi'^2} -1}, \qquad
		\sqrt{1 + \left(\frac{dh}{d\rho}\right)^2} = \frac{2}{\varphi'} \label{eq:dhdrho}.\eeq
	We also note that
	\begin{align}
	\frac{d^2h}{d\rho^2} = \frac{1}{\rho'}\left(\frac{dh}{d\rho}\right)' &= \frac{2e^{-\varphi/2}}{\varphi'} \frac{d}{du}\sqrt{\frac{4}{\varphi'^2} -1} \nonumber \\
    &= -8e^{-\varphi/2}\frac{\varphi''}{\varphi'^3\sqrt{4 - \varphi'^2}} \label{eq:d2hdrho2} .
 \end{align} 
	
	Now substituting Eqs.~\eqref{eq:dhdrho} and \eqref{eq:d2hdrho2} into the principal curvatures (Eq.~\eqref{eq:kappadefn}) gives
 \begin{subequations}
	\begin{align}
		\kappa_1 &= -\frac{\varphi''e^{-\varphi/2}}{\sqrt{4 - \varphi'^2}} \\
		\kappa_2 &= \frac{1}{2}e^{-\varphi/2} \sqrt{4 - \varphi'^2} .
	\end{align}
 \label{eq:kappa}
 \end{subequations}
	Note that in these coordinates, the Gaussian curvature $R$ takes the form
	\beq R = \kappa_1 \kappa_2 = -\frac{1}{2}\varphi'' e^{-\varphi} .\eeq
	
	Upon substitution of Eq.~\eqref{eq:kappa} into $\mathcal F = \mathcal F_Q + \mathcal F_g + \mathcal F_{el}$, we arrive at
    \begin{subequations}
	\begin{align}
		\mathcal F_Q &=  2(K+K')\times\nonumber\\
        &\qquad{}\left[-\pi \sum_m \left(\sigma_m - \frac{1}{2}\sigma_m^2\right)\varphi(u_m) + \frac{1}{8}\int d^2u \, \varphi'^2\right] \label{eq:FQ}\\
		\mathcal F_g &= \int d^2u \, e^{\varphi} [K_\varphi R\varphi + \lambda] \\
		\mathcal F_{el} &= B \int d^2u \, \left(\frac{\varphi''}{\sqrt{4 - \varphi'^2}}+ \frac{1}{2}\sqrt{4 - \varphi'^2}\right)^2 ,\label{eq:el}
	\end{align}
 \label{eq:FTotal}
 \end{subequations}
	completing the task at hand.
	
	We now comment on the effect of $\mathcal F_{el}$ via the principal curvatures $\kappa_1$ and $\kappa_2$. We first note that near a singularity, $\kappa_1$ (the first term in the parentheses of Eq.~\eqref{eq:el}) vanishes whereas $\kappa_2$ (the second term in the parentheses of Eq.~\eqref{eq:el}) is finite. Hence $\kappa_2$ contributes to the coefficient of $\varphi'^2$ in the last term in Eq.~\eqref{eq:FQ}, leading to enchanced charge
	$$ \chi_j = \frac{K+K'}{2K_\varphi + K + K' - B}\left(\sigma_j - \frac{1}{2}\sigma_j^2\right),$$
	which is valid when
	$$ \frac{B}{2K_\varphi + K + K'} < 1 .$$
	
	We now comment on what we must choose for $\lambda$ to keep constant the surface area. Let $\hat {\mathcal F} = \mathcal F - \lambda \int d^2u e^{\varphi} = \mathcal F - \lambda A$. Then
	$$ 0 = \partial_tA = \int d^2u e^{\varphi}\partial_t\varphi = \int d^2u \frac{\delta\mathcal F}{\delta \varphi} = \int d^2u \frac{\delta\hat{\mathcal F}}{\delta \varphi} + \lambda A$$
	from which follows that
	\beq \lambda = -\frac{R_0}{2}\int d^2u \frac{\delta\hat{\mathcal F}}{\delta \varphi} \label{eq:lambda}\eeq
	to fix $A(t)=2/R_0$. In other words, minimizing the free energy while fixing the area is equivalent to minimizing
	$$ \mathcal F = \hat{\mathcal F} + \lambda \left(\int d^2u e^{\varphi} - 2R_0^{-1}\right) ,$$
	where we are treating $\lambda$ as a Lagrange multiplier, given in Eq.~\eqref{eq:lambda}.

 	\begin{figure}[t]
		\centering
	\subfloat[]{\includegraphics[width=.4\linewidth]{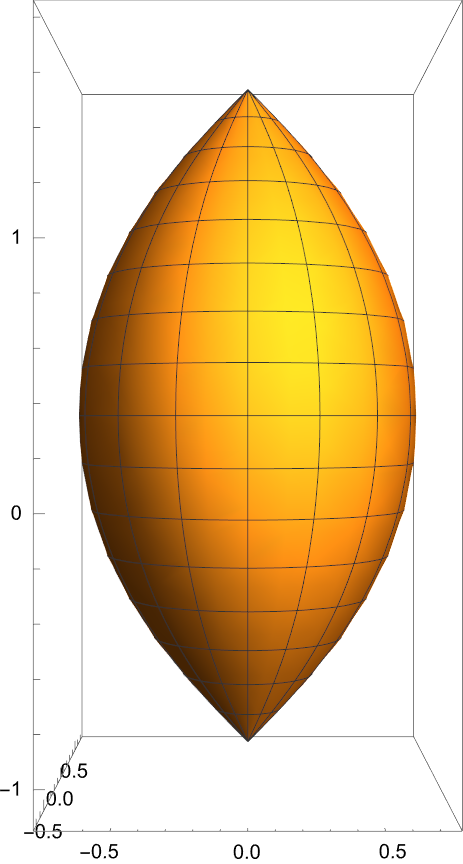}}
  \subfloat[]{\includegraphics[trim=-1cm -1cm 0cm 0cm, width=.55\linewidth]{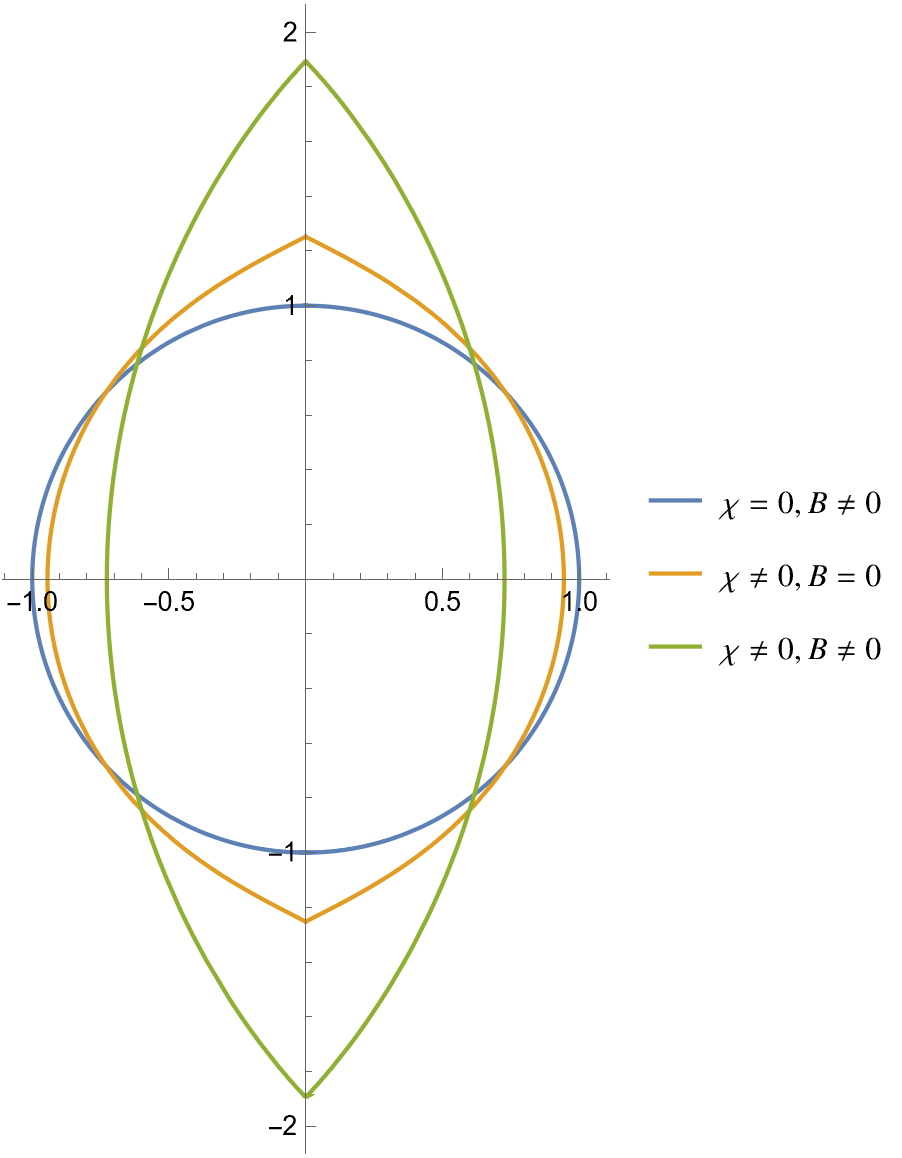}}\\
\subfloat[]{\includegraphics[width=\linewidth]{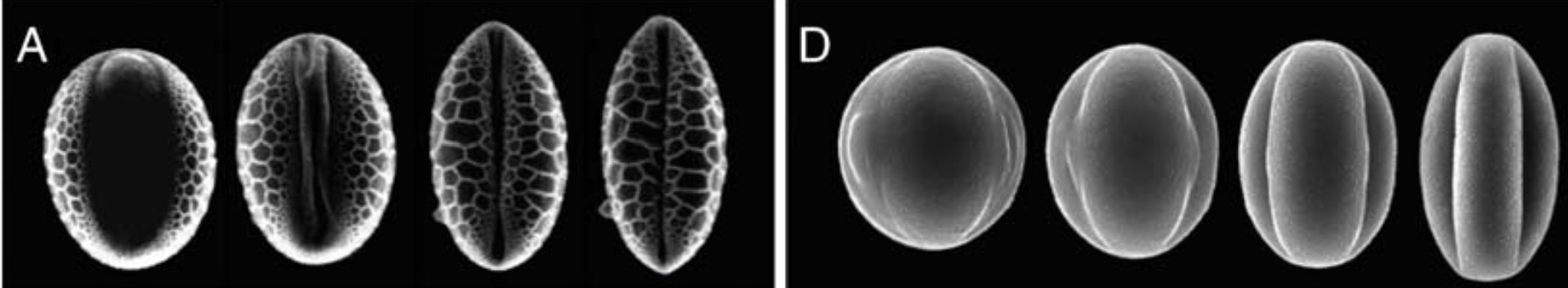}}
		\caption{(a): plot of lemon geometry (Eq.~\eqref{eq:lemon}) for $a=0.7$. (b): plot of long-time solution of Eq.~\eqref{eq:dvarphidtfull}, starting from initial configuration of sphere. Parameters used are $\chi=0.1$ and $\frac{B}{K+K'+2K_\varphi} = 0.6$. (c): figure adapted from Fig. 2 of  Ref.~\cite{katifori2010foldable}. For both lily pollen (left) and euphorbia pollen (right), spherical shapes evolve into lemon shapes.}
		\label{fig:lemon}
	\end{figure}
 
To analyze the dynamics of shaping these surfaces requires the formulation of a gradient flow based on the free energy contributions given by $\mathcal F = \mathcal F_Q + \mathcal F_g + \mathcal F_{el}$ leading to an equation of the form
$$ \partial_t\varphi = -\frac{1}{\sqrt{g}}\frac{\delta \mathcal F}{\delta \varphi} . \label{eq:dvarphidtfull}$$
Here, we do not solve this complicated equation but resort to heuristic arguments to illuminate the basic physics. To gain insight, we consider the initial geometry of the lemon, and study the deformation of this lemon geometry due to the flow. We start with lemon geometry such that when $B=0$, the lemon is a stationary solution. We now turn on the bending term, and consider the effect of $B>0$ for the dynamics. Naively, we would expect the bending energy to flatten the lemon such that the principal curvatures are the same, as in the sphere. We see in Fig.~\ref{fig:lemon} that indeed, near the equator ($u=0$), the geometry does indeed become flattened. However, near the tips, in order to keep the area constant, the tips become more conical.
	
	\subsection{Connecting our results to the shapes of pollen grains}

We now apply our understanding of the previous results to the shape of some spherical shell-like pollen grains that fold into reversible lemon-like shapes when they are dehydrated, and reverse into spherical shells when hydrated~\cite{katifori2010foldable}. We assume polar order, i.e. $p=1$. We note that on the sphere, the low energy configuration involves two $+1$ defects at the north and south poles~\cite{lubensky1992orientational}.
	
 Although, so far we have assumed that  we are in the ordered phase, i.e., $|Q| \neq 0$, we now extend the potential $V[Q]$ to allow $|Q| = 0$, i.e. account for the disordered phase as well. In terms of the humidity $\rho$, and a critical humidity $\rho_c$ where the grain switches from a spherical to a conically folded phase, we let $V[Q]$ takes the form
	$$ V[Q] = \epsilon^{-2}(1 + r |Q|^2)^2 $$
	where $r \propto (\rho - \rho_c)$. For $r> (<) \, 0$, $Q = (\neq) \, 0$. There is thus a 2nd order phase transition at $\rho = \rho_c$, which separates the hydrated and hydrated phases. In the hydrated phase, corresponding to $Q=0$, there are no topological defects, and the pollen grains remain spherical.  In the dehydrated phase, corresponding to $Q \neq 0$, two topological defects at the north and south poles drive the pollen grains to take the shape of a lemon, deformed by the bending energy term, as in Fig.~\ref{fig:lemon}. This allows us to recover the two different geometries (Fig.~\ref{fig:lemon}) shown in ~\cite{katifori2010foldable} and provide an explanation for their origin in terms of the need to have topological defects that drive these shape changes.
	
	\section{Discussion}
	\label{sec:discussion}
	
Our minimal framework for the geometry of curved surfaces with frozen $p$-atic defects driven by relaxational dynamics leads to their diffusive equilibration. In particular, we show that a positive (negative) defect can dynamically generate a cone (hyperbolic cone), and we predict that the half cone angle $\beta$ satisfies $1-\sin\beta \le 1/p (1 - 1/(2p))$. Although we focused primarily on the intrinsic geometry of the surfaces, we showed that for axisymmetric surfaces, where the extrinsic geometry can be deduced entirely by the intrinsic geometry, we can deduce the changes in extrinsic shape as well. For nominally flat surfaces, this leads to a simple intuitive prediction that in the presence of a positive defect, a bump forms with height profile $h(t) \sim \sqrt{t}$ for early times $t$, while for polar order on spheres, we find that the resulting stationary geometry is a deformed lemon. 
	
More generally, we can ask what would happen if the defects were mobile, and moved in response to spatial variations in the geometry, while themselves changing the surface geometry. Over long times, if we have both positive and negative defects, naively we would expect them to annihilate each other. However, if we consider charges of the same sign, as we did for the case of the sphere, there can in principle be a steady state for the defects. For example, Ref.~\cite{lubensky1992orientational} found equilibrium configurations of the $p$-atic defect on a sphere. Here, using this equilibrium configurations as our initial condition for geometric growth, we find that the defects will not move, but the surface will develop conical singularities at the locations of the defects, thus pinning the defects. For example, for the case of $p=1$, we get the lemon configuration. Understanding the varying equilibrium configurations as a function of the number and type of defects is a natural next question to study.
	
While our current study has mainly focused on positive defects, there are other cases where an equilibrium configuration with both positive and negative defects can be attained. For example, on a torus with varying mean and Gaussian curvature, plus-minus defect pairs can nucleate~\cite{bowick2004curvature}. In active systems, we might expect that activity can stabilize a defect configuration of both signs, as for example shown in~\cite{vafa2021active}, and another interesting question is to study these cases further.

With the recently increasing interest in the mathematical and physical study of textiles~\cite{persson2018actuating,yasuda2021mechanical} that are knit or woven from filaments, or active versions thereof, our study suggests  new ways to engineer shape by using $p$-atic defects to generate complex curvature patterns and enhance drapability of the human body, building on ancient empirical approaches that have been long known to artists and artisans.

\acknowledgments{We thank David Nelson and Grace Zhang for valuable discussion of defect dynamics on a cone and Pengfei Guan, Craig Hodgson, Puskar Mondal, Freid Tong, Marc Troyanov, and Shing-Tung Yau for valuable discussions on Ricci flow equations and reconstructing the embedding from the intrinsic metric. We also like to thank Yeonsu Jung for discussions of experimental realizations of the model. This work is partially supported by the Center for Mathematical Sciences and Applications at Harvard University (F. V.), the NSF Simons Center for Mathematical and Statistical Analysis of Biology Award No. 1764269 (L. M.), the Simons Foundation (L. M.), and the Henri Seydoux Fund (L. M.).}
	
\bibliography{refs}
	
\end{document}